# Resynchronization of the Asynchronous Polar CD Ind

Gordon Myers
CBA (San Mateo) and AAVSO, 5 inverness Way, Hillsborough, CA 94010, USA;
gordonmyers@hotmail.com

Joseph Patterson
Department of Astronomy, Columbia University, 550 West 120$^{th}$ Street, New York, NY 10027, USA;
jop@astro.columbia.edu

Enrique de Miguel
CBA (Huelva), Observatorio Astronomico del CIECEM, Parque Dunar Matalascañas, 21760 Almonte, Huelva, Spain;
Departamento de Ciencias Integradas, Facultad de Ciencias Experimentales, Universidad de Huelva, 21071 Huelva, Spain; edmiguel63@gmail.com

Franz-Josef Hambsch
CBA (Mol), Oude Bleken 12, B-2400 Mol, Belgium; hambsch@telenet.be

Berto Monard
CBA (Pretoria), PO Box 281, Calitzdorp 6660, Western Cape, South Africa; bmonard@mweb.co.za

Greg Bolt
CBA (Perth), 295 Camberwarra Drive, Craigie, Western Australia 6025, Australia; gbolt@iinet.net.au

Jennie McCormick
CBA (Pakuranga), Farm Cove Observatory, 2/24 Rapallo Place, Farm Cove, Auckland, New Zealand, 2012; farmcoveobs@xtra.co.nz

Robert Rea
CBA (Nelson), Regent Lane Observatory, 8 Regent Lane, Richmond, Nelson 7020, New Zealand; reamarsh@ihug.co.nz

William Allen
CBA (Blenheim), Vintage Lane Observatory, 83 Vintage Lane, RD 3, Blenheim 7273, New Zealand; whallen@xtra.co.nz

# Abstract

CD Ind is one of only four confirmed asynchronous polars (APs).  APs are strongly magnetic cataclysmic variables of the AM Herculis subclass with the characteristic that their white dwarfs rotate a few per cent out of synchronism with their binary orbit. Theory suggests that nova eruptions disrupt previously synchronized states. Following the eruption, the system is expected to rapidly resynchronize over a timescale of centuries.  The other three asynchronous polars -  V1432 Aql, BY Cam and V1500 Cyg - have resynchronization time estimates ranging from 100 to more than 3500 years, with all but one being less than 1200 years.  We report on the analysis of over 46000 observations of CD Ind taken between 2007 and 2016, combined with previous observations from 1996, and estimate a CD Ind resynchronization time of 6400 ± 800 years.  We also estimate an orbital period of 110.820(1) minutes and a current (2016.4) white dwarf spin period of 109.6564(1) minutes.

Keywords: novae, cataclysmic variables, stars: individual (CD Ind, V1432 Aql, BY Cam, V1500 Cyg),

# 1    Introduction

CD Ind was discovered during the EUVE all-sky survey (EUVE 2115-58.6: Bowyer et al. 1996) and the ROSAT all sky survey (RX J2115-5840: Voges et al. 1996).  Optical spectra (Craig, 1996), spectroscopic observations (Vennes et al. 1996) and optical polarimetry (Schwope et al. 1997) confirmed CD Ind to be a member of the polar subclass of magnetic cataclysmic variables.

In polars the white dwarf's strong magnetic fields (typically > 10 MG) preclude the formation of an accretion disk.  Instead, matter from the secondary star flows directly along the magnetic field lines onto one or both of the white dwarf's magnetic poles.  The strength of the magnetic fields of polars typically synchronizes the spin of the white dwarf with the orbital period of the binary system.

But asynchronous polars (APs) – a small subset of the polars – exhibit orbital periods near but not equal to the white dwarf spin.  Optical polarimetry observations (Ramsay et al., 1999) confirmed CD Ind as a member of this rare subclass. CD Ind has a 1.2 per cent difference between the white dwarf spin and binary orbital period.

Currently there are four confirmed asynchronous polars – V1432 Aql, BY Cam, V1500 Cyg and CD Ind.[1]

---

[1] A fifth star, 1RXS J052430.2+424449 ("Paloma") is sometimes shown as a fifth member of the AP class. While it shows the two-period behavior that is the main credential for the class (Schwarz et al. 2007), its light curve shows many additional periodic signals and its proper classification is still uncertain.

The cause of their asynchronous behavior is theorized to be an eruption on the surface of the white dwarf disrupting a previously synchronized state. V1500 Cyg was the first detected AP showing the asynchronous condition following its 1975 nova eruption. The three other stars plausibly fit this scenario, but are more speculative since the nova event blamed for the asynchronism is only hypothesized, not actually observed. A recent attempt to estimate eruption dates of APs by searching for their nova shell remnants resulted in no shell detections (Pagnotta et al., 2016).

Following the nova disruptive event, it is theorized that resynchronization between the white dwarf and orbital period occurs rapidly (on the order of centuries).[2] Table 1 shows current estimates for V1432 Aql, BY Cam, and V1500 Cyg resynchronization times ($t_{sync}$). The objective for this study is to determine CD Ind's resynchronization time and compare it to the other three asynchronous polars.

**Table 1** – The Asynchronous Polars

| System | $P_{orb}$ (h) | $P_{spin}$ (h) | $(P_{orb}-P_{spin})/P_{orb}$ | $t_{sync}$ in years [source] |
|---|---|---|---|---|
| V1432 Aql | 3.3655 [1] | 3.3751 [2] | -0.0029 | 110[1]; 199[3]; 96.7 [4] |
| BY Cam | 3.3544 [1] | 3.3222 [2] | 0.0096 | 1200[5]; 1107[1]; ≥3500[6]; 250[7] |
| V1500 Cyg | 3.3507 [1] | 3.2917 [2] | 0.0176 | 185[5]; 150[1]; 150-290 [7]; 39[9]; 170[10] |
| CD Ind | 1.8467 [1] | 1.8258 [2] | 0.0113 | 6400 [8] |

References: Pagnotta & Zurek (2016); [1] Campbell & Schwope (1999); [2] Ramsay et al. (1999); [3] Staubert et al. (2003); [4] Andronov & Baklanov (2007); [5] Piirola et al. (1994); [6] Honeycutt & Kafka (2005); [7] Pavlenko et al. (2013); [8] this paper; [9] Harrison & Campbell (2016); [10] Schmidt et. al (1995); [10] Schmidt et. al. (1995)

## 2 Observations

More than 46000 photometric observations taken over 1224 hours on 285 nights were submitted to the Center for Backyard Astrophysics (CBA) (www.cbastro.org) by eight observers between 2007 and 2016. Table 2 shows the spread of observations across those 10 years.

**Table 2.** Summary of Observations

| Year | No. of Runs | No. of Observations | Duration of Observing Sessions (h) | Approximate Number of Spin Cycles Observed |
|---|---|---|---|---|
| 2007 | 31 | 8631 | 153 | 83 |

---

[2] This point is disputed by Harrison & Campbell (2016) who interpreted their 2014 observations to indicate V1500 Cyg has already synchronized – a mere 39 years after the nova. This point is important since V1500 Cyg is the only AP for which we know the actual date of the nova event. But CBA's extensive photometry in 2014 shows an obvious detection of the spin period, free from aliases and shorter than P_orb by 175 ± 1 s. This is consistent with the other photometrically derived timescales for synchronization in Table 1, as well as the timescale of 170 yr derived by Schmidt et. al. (1995) from polarimetry. These CBA observations will be written up separately.

| | | | |
|---|---|---|---|
| 2008 | 20 | 10273 | 103 | 56 |
| 2011 | 53 | 13467 | 270 | 148 |
| 2013 | 9 | 879 | 27 | 14 |
| 2014 | 52 | 1923 | 163 | 89 |
| 2015 | 76 | 7144 | 295 | 161 |
| 2016 | 44 | 3771 | 213 | 116 |
| Totals | 285 | 46088 | 1224 | 667 |

The CBA is a global network of small telescopes dedicated to photometry of cataclysmic variables. The observations were taken with telescopes located in South Africa, Chile, Australia and New Zealand. This provides a good longitudinal spread significantly reducing aliases. Table 3 describes the telescopes used by the CBA observers.

**Table 3.** CBA telescopes used to observe CD Ind from 2007 to 2016.

| Observer | Observatory Location | Telescope | CCD | Number of Measured Maxima |
|---|---|---|---|---|
| Hambsch | San Pedro de Atacama, Chile | Orion Optics ODK 400 mm | FLI w/Kodak 16803 | 64 |
| Myers | Coonabarabran, Australia | Planewave CDK 410 mm | FLI PL4710 | 48 |
| Rea | Nelson, New Zealand | 2007/2008 - Celestron C11 | SBIG ST-402 | 5 |
| | | 2007/2008 - Celestron C14 | SBIG ST-9 | 15 |
| | | 2011 - CDK20 | SBIG ST-9 | 5 |
| McCormick | Auckland, New Zealand | Meade LX200R - 360 mm | SBIG ST-8XME | 3 |
| Monard | Pretoria, South Africa | 2007/2008-Meade RCX400 - 300 mm | SBIG ST-7XME | 18 |
| | Calitzdorp, South Africa | 2010- Meade RCX400-350 mm | SBIG ST-8XME | 3 |
| Bolt | Perth, Australia | Meade LX200 - 254 mm | SBIG ST-7 | 3 |
| Allen | Blenheim, New Zealand | Classical Cassegrain - 400 mm | SBIG STL-1001E | 1 |

Observations of CD Ind used exposure times between 30 and 120 sec and either no filter or Clear filters with a UV cutoff. All images were dark subtracted and flat fielded. Timings were adjusted to HJD and observations from different observers were adjusted to yield a common instrumental magnitude, with effective wavelength near 6000 A ("pink"). Observers used differential photometry using AAVSO and UCAC4 comp magnitudes. Overlap of observing runs between observers were used to develop constants for adjusting the magnitudes to a common baseline.

## 3 Results

## 3.1 White Dwarf Spin
## 3.1.1 Analysis based on CBA 2007-2016 Observations

Obtaining an accurate value of the spin period requires careful timing of a specific feature observed in each spin cycle. But the hump profiles in CD Ind vary considerably - often changing from night to night, and complicating the task of choosing a consistent marker.

In a fully synchronized polar the accretion flow from the secondary follows a ballistic trajectory until captured by the magnetic field of the primary where it is directed above or below the orbital plane onto the surface of the white dwarf (WD). The relative position of the two stars remains fixed because the WD spin and binary orbit period are locked. A more complex flow is expected to occur in APs since the accretion region on the surface of the white dwarf is not fixed, but changes periodically on the beat cycle (approximately 7.3 days for CD Ind as discussed and calculated in section 3.3).

Like nearly all magnetic CV's, CD Ind sometimes dives into states of very low accretion. The exact cause is not known. This was observed twice during our observations (HJD 2455830.4 and 2457225.2).

The variation in the profile makes it difficult to obtain precise timing measurements of a consistent feature in each cycle. Figure 1 shows three examples of the changing profile probably caused by the ever-changing angle between the two magnetic fields (the white dwarf and the secondary).

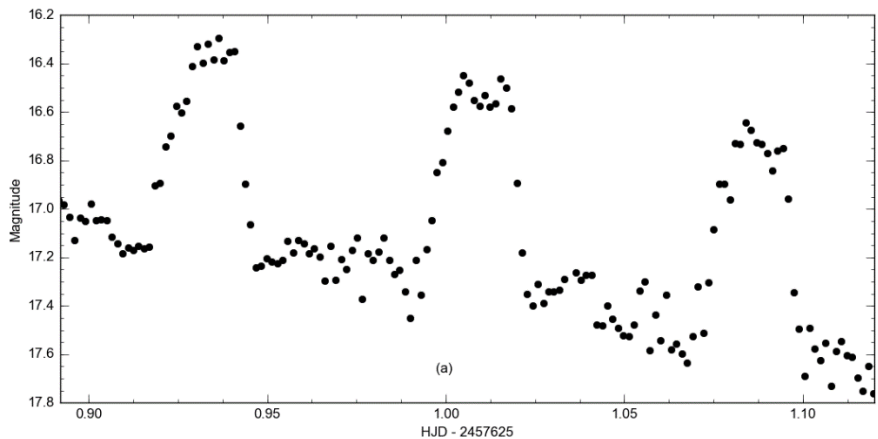

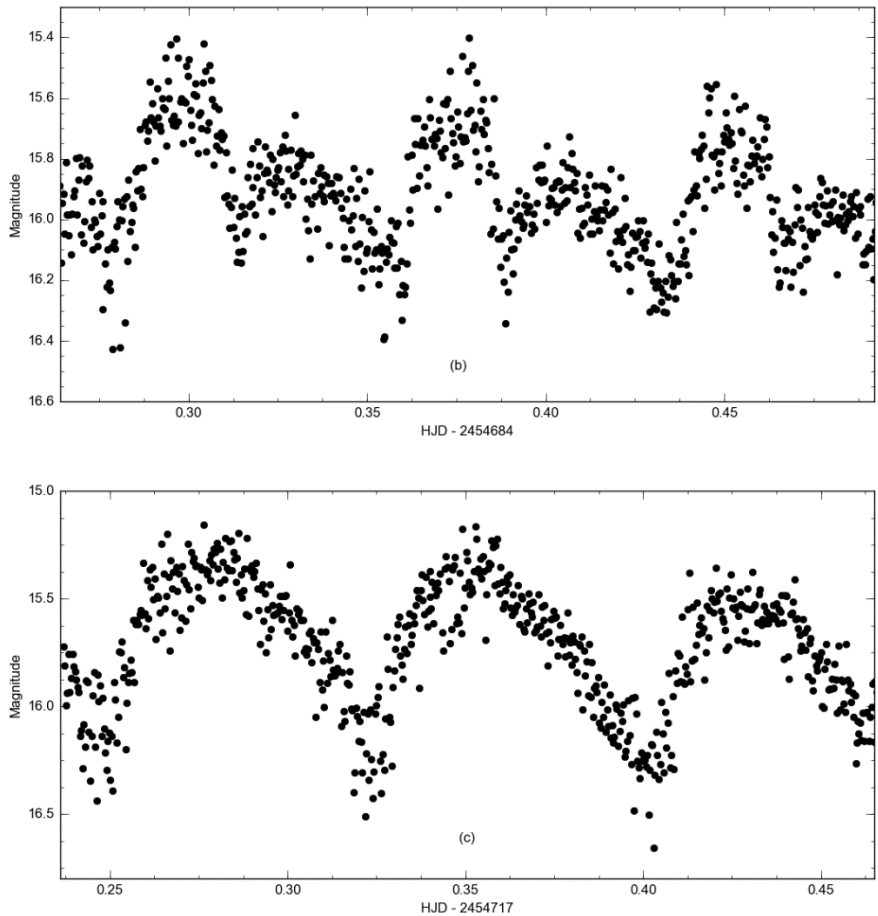

**Figure 1** Several nightly light curves of CD Ind, illustrating the variety. We interpret type (a) as signifying accretion onto one magnetic pole, and type (b) as possibly signifying two pole accretion. Type (c) probably arises from one main pole accreting with a changing perspective angle during the white dwarf spin cycle. With this variety we choose only to use type (a) for our period study.

In order to improve the accuracy of the timing of individual maxima, only cycles with light curves clearly showing a profile similar to Figure 1(a) were used. Of the approximate 667 spin cycles observed, 165 had this distinctive profile. Even with these light curves there was no clear marker for precise timings. Many humps exhibit virtually flat profiles near their peak. Therefore, times of maximum light were determined by fitting a quadratic curve to the observed humps and recording the time of maximum light. Figure 2 shows the quadratic fits to three of the peaks observed using different scopes which used different image exposure times.

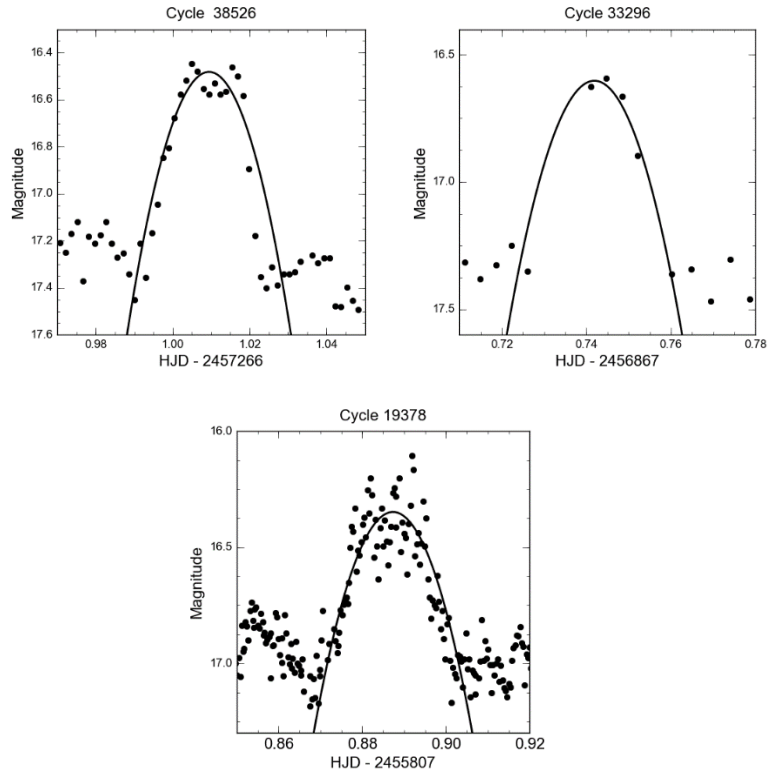

**Figure 2**. For all profiles of type (a) we determine the time of maximum light by fitting a quadratic. The three plots show different image exposure times used by different observers – 60, 120 and 30 seconds.

After we measured the timings on the 165 selected humps we generated ephemerides using linear and quadratic regressions on the times of maximum light as a function of associated cycle counts.

The linear regression of the measured times of maximum light vs. cycle count yields the following ephemeris with an O-C standard deviation of .0038 d:

HJD of mid-humps = 2454332.2616(38) + 0.07614973(9) E                     (1)

where E is the cycle number. The O-C residuals using equation (1) are shown in Figure 3. A quadratic regression to these residuals yields the curve shown indicating a changing white dwarf spin period.

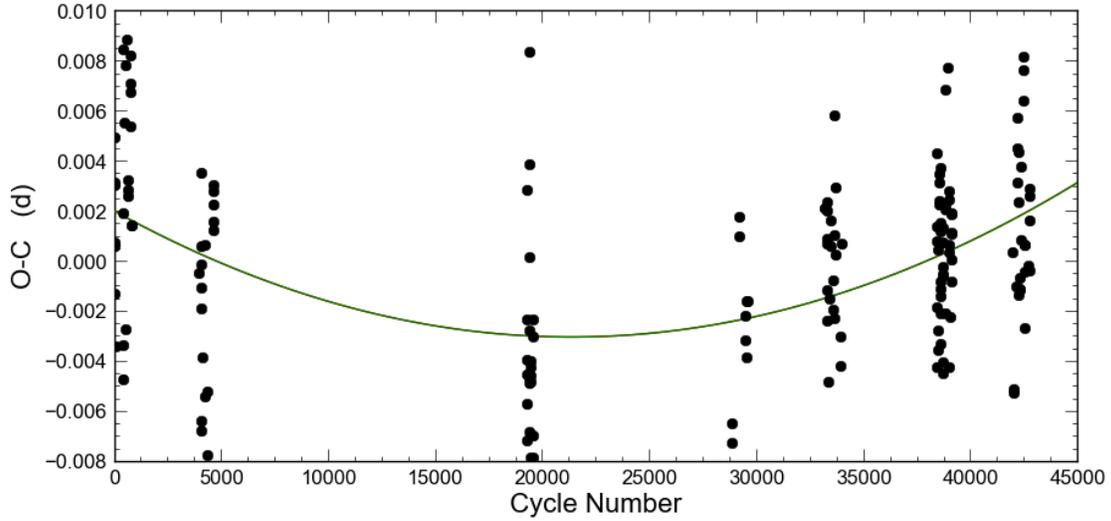

**Figure 3**. O-C diagram of mid-hump times, relative to a test period of 0.07614973 days (equation (1)). The quadratic fit yields $dP_{spin}/dt = 2.9(4) \times 10^{-10}$

The quadratic regression of the measured times of maximum light vs. cycle count yields the following ephemeris with an O-C standard deviation of 0.0034 d :

HJD of mid-humps = 2454332.2637(34) + 0.07614926(8) E + $1.10(18) \times 10^{-11}$ $E^2$   (2)

### 3.1.2 Incorporation of Schwope et al. data from 1996

In addition to the 165 timings measured using the CBA data, four timings of mid-hump peaks observed in 1996 were available from Schwope et al. (1997). Because of uncertainty about exactly how the measurements were made – and the fact there is an 11-year gap with the CBA data, analysis was first performed on only the CBA data. Then the effect of including Schwope's additional four earlier measurements was evaluated.

Schwope et al. 1997 contains four mid-hump timings "…derived for far-pole accretion" so it appears measurements were made on light curve profiles similar to those used in analyzing CBA data.

A quadratic regression to the combined data yields the following ephemeris with an O-C standard deviation of 0.0034 d:

HJD of mid-humps = 2454332.2641(34) + 0.07614917(8) E + $1.32(18) \times 10^{-11}$ $E^2$   (3)

The O-C residuals incorporating the Schwope data are shown in Figure 4. From the second-order coefficient in equation (3) we infer a value of $dP_{spin}/dt = 3.5(4) \times 10^{-10}$. Because of the consistency of the results when incorporating the Schwope data and the fact that with these data we have a 20-year observation baseline, equation (3) is adopted as the best ephemeris. Based on this ephemeris the white dwarf spin period increases from 109.6528(1) to 109.6564(1) min across the 20 years of observation. This period is consistent with the lower of the two estimates (109.84 or 109.65 minutes) from Schwope et al. (1997) which were based on photometric measurements.

In order to be conservative with our error estimates we used the 42752 cycle count from the CBA data rather than the full 95251 cycles starting with Schwope's 1996 observations.

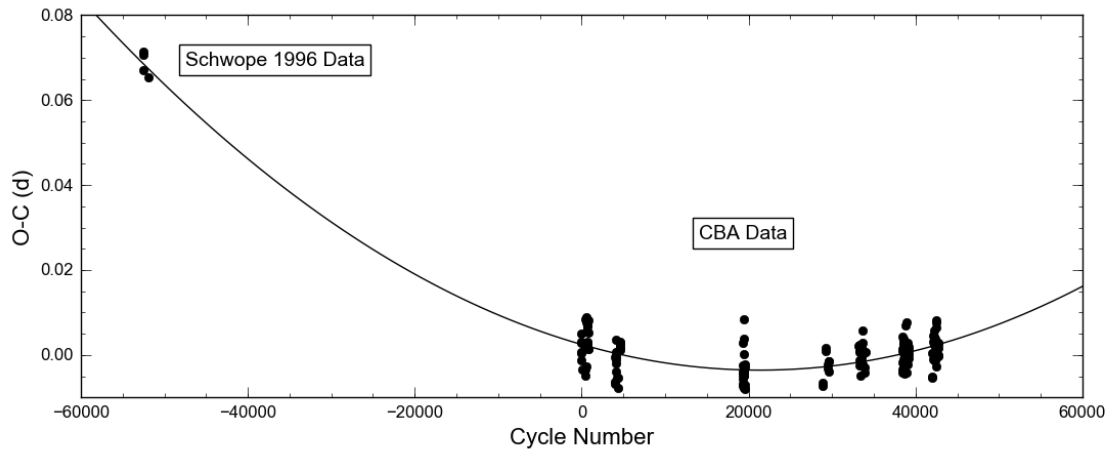

Figure 4. O-C diagram of mid-hump times relative to a test period of 0.07614973 days  The quadratic fit yields $dP_{spin}/dt = 3.5(4) \times 10^{-10}$

## 3.2 Orbital Period

We analyzed all the CBA data using the Period04 package (Lenz and Breger, 2005) based on the discrete Fourier transform method. Based on existing information we limited the period search to the interval 12.5 to 13.5 cycles/d to include both the orbital and spin periods. We show the corresponding power spectrum in Figure 6. The spectrum is dominated by a peak centered at 13.1320(14) cycles/d, a value fully consistent with our previous estimate of the spin period.

A peak at a frequency of 12.994939(5) cycles/d, or a periodicity of 110.82005(4) min, is also apparent in the spectrum. This is similar to previous reports of the orbital period of 110.889 minutes from Ramsay et al. (1999) based on polarimetry data and 110.75(6) minutes from Vennes et al. (1996) based on spectroscopy data. Thus we interpret this detection as a footprint of the orbital motion. The error estimates were obtained using Period04's Monte Carlo simulation.

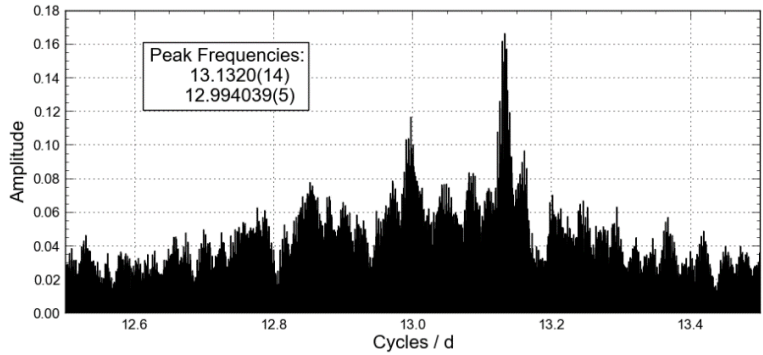

**Figure 5**. Power Spectrum – the two dominant peaks correlate to the orbital period (12.99 c/d) and the white dwarf spin (13.13 c/d)

## 3.3 Beat Period

Based on the WD 2015 spin frequency ($f_{spin}$) of 13.1324 c/d and orbital frequency ($f_{orbit}$) of 12.9949 c/d, the 2015 beat frequency of 0.137 c/d which equates to a period of 7.299 d. Since the beat period represents the time taken by the secondary to orbit the white dwarf in the frame of the rotating white dwarf, a clear beat period profile was expected. For unknown reasons the strength of the appearance of the beat period varies considerably during different years. The observing year with the most runs – 2015 with 76 runs - shows the most distinctive beat period pattern. Figure 6 shows the light curve folded on the beat period of 7.299 d

The beat period assumes the white dwarf spin is prograde (i.e. the beat frequency is $f_{spin}$ - $f_{orbit}$ = 0.137 c/d). If the motion were retrograde, the beat frequency would be $f_{spin}$ + $f_{orbit}$ = 26.127 c/d. Spectral signals were detected at both frequencies with the 0.137 c/d signal having a power amplitude of 0.35 compared to a power amplitude of 0.078 for the 26.127 c/d signal. The significantly higher strength of the 0.137 c/d signal supports the assumption of prograde motion – consistent with the current theory of AP formation.

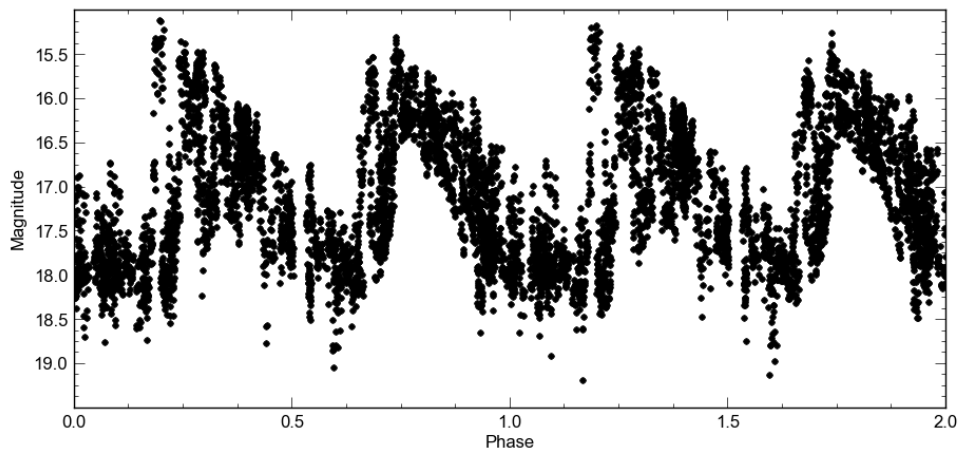

**Figure 6.** 2015 Folded Light Curve based on a beat period of 7.299 d

## 4 Conclusion - Resynchronization Projection

A resynchronization time estimate was made by dividing the current difference between the WD spin and orbital spin periods ($P_{spin}$ and $P_{orb}$) by the white dwarf spin change rate $dP_{spin}/dt = 3.5(4) \times 10^{-10}$. This assumes the spin change rate is constant over time and yields a resynchronization time of 6400 +/- 800 years.

This predicted resynchronization time is longer than estimates for the other three APs shown in Table 1. The exact reason is not known. One possibility is that the resynchronization rates may not be constant; they may vary over time – e.g. they could decay exponentially as the systems approach synchronism.

Another potential cause relates to the strength of the CD Ind's relatively low magnetic field strength (for a polar). Polars field strength is typically 7 - 230 MG (Ferranio, de Martino & Gänsicke, 2015). Estimates for CD Ind's field are 11±2 MG (Schwope et al. 1992) and <20 MG (Vennes et al. 1996). Campbell & Schwope (1999) indicate the resynchronization time is inversely proportional to the square of the magnetic field strength. This may be a contributing factor but additional research is required.

Separate from the analysis of the resynchronization time estimate, the question of whether CD Ind's WD spin is prograde or retrograde was examined. The Fourier power spectrum at the two potential interaction frequencies ($f_{spin} + f_{orb}$ and $f_{spin} - f_{orb}$) was examined. Based on the strength of the signals, CD Ind exhibits prograde motion which is consistent with the current theory of AP evolution.

## 5 Acknowledgements

We thank the National Science Foundation for support of this research (AST-1211129 and AST-1615456), and NASA through HST-GO-13630. Enrique de Miguel acknowledges financial support from the Ministerio de Educacion, Cultura y Deporte (Spain) under the Mobility Program Salvador de Madariaga (PRX15/00521).

This research made use of Astropy, a community-developed core Python package for Astronomy (Astropy Collaboration, 2013).## 6 References

Andronov I. L., Baklanov A. V., 2007, Astrophysics, 50, 105
Bianchini A., Saygac T., Orio M., della Valle M., Williams R., 2012, A&A, 539, A94
Bowyer S., Lampton M., Lewis J., Wu X., Jelinsky P., Malina R. F., 1996, ApJS, 102, 129
Campbell C. G., Schwope A. D., 1999, A&A, 343, 13211

Cropper, Mark (1990-12-01). "The polars". Space Science Reviews 54 (3-4): 195–295.
Ferrario L., de Martino D., Gänsicke B. T., 2015, Space Science Rev., 191, 111
Harrison T., Campbell R., 2016, MNRAS, 459, 4161Honeycutt R. K., Kafka S., 2005, MNRAS, 364, 917
Lenz P., Breger M., 2005, Commun. Asteroseismol., 146, 53
Lipkin Y. M., Leibowitz E. M., 2008, MNRAS, 387, 289
Pagnotta A., Zurek D., 2016, MNRAS, 458, 1833
Pavlenko E., Andreev M., Babiba Y., Malanushenko V., 2013, in Krzesiński J., Stachowski G., Moskalik P., Bajan K., eds. ASP Conf. Ser. Vol. 469, 18[th] European White Dwarf Workshop., Astron. Soc. Of the Pac., San Francisco, CA, p.343
Piirola V., Coyne G. V., Takalo S. J., Takalo L., Larsson S., Vilhu O., 1994, A&A, 283, 163
Ramsay G., Buckley D., Cropper M., Harrop-Allin M. K., 1999, MNRAS, 303, 96
Ramsay G., Potter S., Cropper M., Buckley D., Harrop-Allon M. K., 2000, MNRAS, 316, 225
Schmidt G., Liebert J., Stockman H., 1995, ApJ, 441, 414
Schwarz R., Schwope A.D., Staude A., Rau A., Hasinger G., Urrutia T., Motch C.,2007, A&A, 473, 511
Schwope A. D., Buckley D. A. H., O'Donoghue D., Hasinger G., Trümper J., Voges W., 1997, A&A, 326, 195
Staubert R., Friedrich S., Pottschmidt K., Benlloch S., Schuh S. L., Knoll P., Splittgerber E., Rothschild R., 2003, A&A, 407, 987
Vennes S., Wickramasinghe D., Thorstensen J., Christian D., Bessell M., 1996, AJ, 112, 2254
Voges W., et al., 1996,, IAU Circ. 6420

**Table 4** – Cycle Numbers and HJD Observed Times of mid-humps

| Cycle Number | HJD | Cycle Number | HJD | Cycle Number | HJD | Cycle Number | HJD |
|---|---|---|---|---|---|---|---|
| -52499 | 2450334.54400 | 4638 | 2454685.44635 | 33479 | 2456881.68010 | 38932 | 2457296.93072 |
| -52474 | 2450336.45130 | 4639 | 2454685.52325 | 33492 | 2456882.66905 | 38987 | 2457301.10695 |
| -52473 | 2450336.52820 | 4640 | 2454685.59918 | 33571 | 2456888.68350 | 38998 | 2457301.94948 |
| -51907 | 2450379.62310 | 19270 | 2455799.66127 | 33584 | 2456889.67229 | 38999 | 2457302.02535 |
| 0 | 2454332.26465 | 19272 | 2455799.81209 | 33624 | 2456892.72601 | 39000 | 2457302.10395 |
| 1 | 2454332.33648 | 19283 | 2455800.65296 | 33663 | 2456895.68774 | 39013 | 2457303.09357 |
| 2 | 2454332.41465 | 19284 | 2455800.73070 | 33676 | 2456896.68103 | 39090 | 2457308.95241 |
| 3 | 2454332.49067 | 19285 | 2455800.80469 | 33689 | 2456897.67021 | 39103 | 2457309.94566 |
| 4 | 2454332.56936 | 19298 | 2455801.80200 | 33715 | 2456899.65275 | 39104 | 2457310.02258 |
| 8 | 2454332.87578 | 19376 | 2455807.73201 | 33951 | 2456917.61816 | 39116 | 2457310.93370 |
| 73 | 2454337.81716 | 19377 | 2455807.81010 | 33952 | 2456917.69311 | 39117 | 2457311.01075 |
| 405 | 2454363.11072 | 19378 | 2455807.88645 | 33978 | 2456919.67791 | 39118 | 2457311.08874 |
| 417 | 2454364.01799 | 19380 | 2455808.04064 | 38409 | 2457257.09482 | 39129 | 2457311.92560 |
| 448 | 2454366.37335 | 19388 | 2455808.65278 | 38410 | 2457257.16857 | 41956 | 2457527.20009 |
| 449 | 2454366.44815 | 19391 | 2455808.88493 | 38429 | 2457258.62104 | 42034 | 2457533.13419 |
| 508 | 2454370.95123 | 19392 | 2455808.96558 | 38430 | 2457258.69661 | 42036 | 2457533.28663 |
| 521 | 2454371.94349 | 19462 | 2455814.27987 | 38433 | 2457258.92859 | 42161 | 2457542.80942 |
| 522 | 2454372.00906 | 19467 | 2455814.66361 | 38499 | 2457263.94660 | 42173 | 2457543.72877 |
| 587 | 2454376.97035 | 19468 | 2455814.74002 | 38516 | 2457265.24191 | 42174 | 2457543.80353 |
| 652 | 2454381.91409 | 19473 | 2455815.12111 | 38521 | 2457265.62589 | 42188 | 2457544.87224 |
| 654 | 2454382.06614 | 19474 | 2455815.19751 | 38525 | 2457265.93229 | 42240 | 2457548.82865 |
| 655 | 2454382.14292 | 19556 | 2455821.43878 | 38526 | 2457266.00859 | 42241 | 2457548.90110 |
| 743 | 2454388.84762 | 19558 | 2455821.59021 | 38527 | 2457266.08546 | 42280 | 2457551.87661 |
| 744 | 2454388.92414 | 19571 | 2455822.58571 | 38528 | 2457266.16198 | 42331 | 2457555.75478 |
| 745 | 2454389.00139 | 19573 | 2455822.73729 | 38607 | 2457272.17220 | 42332 | 2457555.83088 |
| 783 | 2454391.89223 | 28889 | 2456532.14395 | 38608 | 2457272.24716 | 42333 | 2457555.90754 |
| 849 | 2454396.91419 | 28890 | 2456532.22090 | 38613 | 2457272.63011 | 42383 | 2457559.71656 |
| 850 | 2454396.99031 | 29230 | 2456558.12003 | 38619 | 2457273.08930 | 42385 | 2457559.87177 |
| 3965 | 2454634.19483 | 29231 | 2456558.19544 | 38620 | 2457273.16579 | 42464 | 2457565.89022 |
| 4048 | 2454640.50936 | 29478 | 2456577.00024 | 38621 | 2457273.24414 | 42475 | 2457566.72911 |
| 4049 | 2454640.58510 | 29498 | 2456578.52424 | 38626 | 2457273.61977 | 42476 | 2457566.80579 |
| 4059 | 2454641.35151 | 29569 | 2456583.92921 | 38627 | 2457273.69649 | 42517 | 2457569.91709 |
| 4060 | 2454641.42846 | 29570 | 2456584.00761 | 38697 | 2457279.02329 | 42528 | 2457570.75806 |
| 4061 | 2454641.50556 | 29583 | 2456584.99754 | 38699 | 2457279.17603 | 42529 | 2457570.83311 |
| 4062 | 2454641.58245 | 33204 | 2456860.73944 | 38709 | 2457279.94133 | 42530 | 2457570.91037 |
| 4070 | 2454642.19454 | 33282 | 2456866.67463 | 38710 | 2457280.01849 | 42729 | 2457586.06334 |
| 4148 | 2454648.12686 | 33295 | 2456867.66762 | 38718 | 2457280.62825 | 42739 | 2457586.82663 |
| 4241 | 2454655.20721 | 33296 | 2456867.74192 | 38719 | 2457280.70256 | 42740 | 2457586.90079 |
| 4254 | 2454656.20321 | 33309 | 2456868.73538 | 38725 | 2457281.15941 | 42742 | 2457587.05633 |
| 4332 | 2454662.13706 | 33321 | 2456869.64886 | 38810 | 2457287.63479 | 42752 | 2457587.81755 |
| 4333 | 2454662.21066 | 33322 | 2456869.72389 | 38815 | 2457288.01137 | | |
| 4636 | 2454685.29301 | 33374 | 2456873.67794 | 38827 | 2457288.92950 | | |
| 4637 | 2454685.36950 | 33400 | 2456875.66117 | 38842 | 2457290.07636 | | |